\documentclass[aps,twocolumn,showpacs,preprintnumbers,nofootinbib,prd,10pt,superscriptaddress]{revtex4-1}

\makeatletter
\def\l@subsubsection#1#2{}
\def\l@subsubsubsection#1#2{}
\makeatother

\usepackage{graphicx,amssymb,amsmath,amsthm,amsfonts,epsfig,epsf,fixmath}
\usepackage{comment}
\usepackage{mathrsfs}
\usepackage[usenames]{color}
\usepackage{epstopdf}
\usepackage{aas_macros}
\usepackage{bm}
\usepackage{dcolumn}
\usepackage{latexsym}
\usepackage{rotating}
\usepackage{longtable}

\usepackage{enumerate}
\usepackage{tensor,multirow}
\usepackage{url}
\usepackage{float}
\usepackage{array}
\newcolumntype{C}[1]{>{\centering\let\newline\\\arraybackslash\hspace{0pt}}m{#1}}

\usepackage[dvipsnames]{xcolor}
\usepackage[unicode]{hyperref}
\hypersetup{colorlinks=true, citecolor=MidnightBlue,
            linkcolor=MidnightBlue, urlcolor=MidnightBlue, linktocpage=true}
            
\usepackage{tikz}

\newcommand{\be}{\begin{equation}}
\newcommand{\eeq}{\end{equation}}
\newcommand{\ba}{\begin{align}}
\newcommand{\ea}{\end{align}}

\newcommand{\GSSI}{Gran Sasso Science Institute (GSSI), I-67100 L’Aquila, Italy}
\newcommand{\GranSasso}{INFN, Laboratori Nazionali del Gran Sasso, I-67100 Assergi, Italy}
\newcommand{\jhu}{William H.\ Miller III Department of Physics and Astronomy, Johns Hopkins University, \\ 3400 N. Charles Street, Baltimore, Maryland, 21218, USA}

\begin{document}

\title{Extreme mass ratio inspirals in dark matter halos: \\ Dynamics and distinguishability of halo models}

\author{Sara Gliorio}
\email{sara.gliorio@gssi.it}
\address{\GSSI}
\address{\GranSasso}

\author{Emanuele Berti}
\email{berti@jhu.edu}
\address{\jhu}

\author{Andrea Maselli}
\email{andrea.maselli@gssi.it}
\address{\GSSI}
\address{\GranSasso}

\author{Nicholas Speeney}
\email{nspeene1@jhu.edu}
\address{\jhu}

\begin{abstract} 
  The gravitational wave (GW) signals from extreme mass-ratio inspirals (EMRIs), a key target for the Laser Interferometer Space Antenna (LISA), will be affected in the presence of dark matter (DM) halos. In this paper we explore whether the effects of DM are detectable by LISA within a fully relativistic framework. We model the massive EMRI component as a nonrotating black hole (BH) surrounded by a DM halo. We compute axial and polar GW fluxes for circular orbits at linear order in the mass ratio for DM density profiles with varying mass and compactness. By comparing the phase evolution with vacuum systems, we find that DM halos can induce dephasings of tens to hundreds of radians over a one-year observation period.  We demonstrate that even highly diluted DM distributions can significantly affect the emitted waveforms, and that the resulting GW signals can usually be distinguished from each other. While it is important to generalize these findings to more generic orbits and to spinning BHs, our results suggest that LISA could not only reveal the presence of DM halos, but also discriminate between different halo models.
\end{abstract}

\maketitle

%
\section{Introduction}

A wide range of astrophysical observations indicate that dark matter (DM) is the most abundant form of matter in the Universe~\cite{Freese:2008cz,Navarro:1995iw,Clowe:2006eq,Bertone:2004pz}. Understanding the nature of DM is one of the main challenges in modern science, and extensive research efforts are devoted to the exploration of possible interactions between the dark sector and the Standard Model~\cite{Kahlhoefer:2017dnp,PerezdelosHeros:2020qyt}.

In parallel, gravitational wave (GW) observations from coalescing compact objects have opened a novel avenue for probing DM properties and, more generally, the environments in which binaries evolve. The presence of nonvacuum spacetimes can alter both the generation and propagation of GW signals, leaving distinctive imprints on their waveforms. These effects depend on the characteristics of the surrounding environment, which may include baryonic matter in accretion disks, clouds of new fundamental fields, or DM halos~\cite{Yunes:2011ws,DeLuca:2022xlz,Sberna:2022qbn,Speri:2022upm,Kavanagh:2020cfn,Macedo:2013qea,Eda:2013gg,Eda:2014kra,Babak:2006uv,Destounis:2021mqv,Destounis:2022obl,Vicente:2022ivh,Traykova:2021dua,Sadeghian:2013laa,Speeney:2022ryg,Boskovic:2018rub,Annulli:2020lyc,Coogan:2021uqv,Cole:2022yzw,Cardoso:2021wlq,Cardoso:2022whc,Figueiredo:2023gas,Baumann:2021fkf,Duque:2023seg,Barsanti:2022vvl,Duque:2024mfw,Brito:2023pyl}.

In general, the DM densities required to induce changes in GW signals that may be observable with current or near-future detectors are expected to be several orders of magnitude higher than the average galactic value (in the solar neighborhood, this average density is approximately
$\sim 0.01M_\odot/{\rm pc}^3 \sim 0.1{\rm GeV}/{\rm cm}^3$).
Probing DM through coalescing binaries thus requires a physical mechanism capable of enhancing DM density around isolated compact objects.

A particularly promising scenario involves the formation of DM spikes around black holes (BHs). Both Newtonian and relativistic calculations predict that particle accretion can lead to the development of overdensities, which redistribute to form a cusp. The specific characteristics of these cusps depend on the initial DM profile, with their spatial extent determined by the BH mass~\cite{Gondolo:1999ef,Sadeghian:2013laa, Konoplya_2022}.

The persistence of these overdensities during binary coalescence remains an open question~\cite{Ullio:2001fb}. Numerical simulations of wave-like DM suggest that overdensities can survive mergers involving comparable-mass binaries, influencing their dynamical evolution~\cite{Bamber:2022pbs,Zhang:2022rex,Choudhary:2020pxy,Schive:2014dra,Aurrekoetxea:2023jwk}. In contrast, for models involving heavy DM particles, systems with nearly equal masses are expected to deplete DM spikes, leading to their dissipation during violent or repeated mergers~\cite{Kavanagh:2018ggo,Bertone:2004pz,Bertone:2005hw}. However, DM overdensities play a crucial role in the dynamics of asymmetric binaries, such as extreme and intermediate mass ratio inspirals (EMRIs/IMRIs).

In particular, EMRIs are promising sources to probe environmental effects through GW observations~\cite{Cardoso:2019rou}.
EMRIs consist of a stellar-mass object orbiting a supermassive BH, with mass ratios of $q=m_p/M_{\rm BH} \sim \mathcal{O}(10^{-5}-10^{-8})$, where $m_p$ and $M_{\rm BH}$ denote the masses of the smaller and larger components, respectively. These systems can spend $\sim 10^5$ orbital cycles in the low-frequency regime accessible to LISA~\cite{LISA:2017pwj}, enabling exquisite measurements of source parameters.

Modeling EMRIs in vacuum general relativity (GR) is already a formidable challenge, which is still incomplete after three decades of impressive theoretical developmens~\cite{Barack:2018yvs}. Extending these calculations to include environmental effects is even more complex, primarily due to two key difficulties: (i) the lack of relativistic solutions describing BHs embedded within matter distributions, and (ii) the intricate couplings between metric and matter variables. As a result, most studies investigating environmental effects on EMRIs rely on approximations, such as post-Newtonian methods or geodesic models that neglect dissipative effects~\cite{Kocsis:2011dr,Kavanagh:2020cfn,Coogan:2021uqv,Cole:2022yzw,Tomaselli:2023ysb,Berezhiani:2023vlo, Rahman_2024}.

To address these limitations, a recent research program introduced a relativistic framework based on the Einstein cluster prescription to model the spacetime of BHs surrounded by spherically symmetric density distributions~\cite{Cardoso:2021wlq,Cardoso:2022whc}. This approach was used to analyze coupled gravitational and fluid perturbations induced by the smaller component of an EMRI when the central BH is embedded in a DM halo. The analysis focused on the Hernquist distribution~\cite{1990ApJ...356..359H}, which allows for closed-form analytic solutions for the background metric. Further advancements extended this formalism by developing a fully numerical framework capable of handling both axial and polar perturbations for general matter distributions surrounding the central BH~\cite{Figueiredo:2023gas,Speeney:2024mas}.

In this paper, we extend this research program by investigating the relativistic evolution of EMRIs in a DM halo. We examine how the surrounding matter affects both the adiabatic GW emission and the orbital dynamics for various families of DM density profiles. We quantify deviations from the vacuum case and assess their detectability by LISA using different figures of merit. Our results demonstrate that EMRI observations across a broad mass range can provide strong evidence for the presence of DM spikes near supermassive BHs. Furthermore, we show that the imprints left on GW waveforms by different DM distributions are significant enough to be distinguishable from one another.
Unless stated otherwise, throughout the paper we use geometrical units ($G=c=1$).

%
\section{Theoretical formalism}
%

In this section we review the formalism developed in Refs.~\cite{Figueiredo:2023gas,Speeney:2024mas} to compute axial and polar perturbations sourced by a massive particle in circular motion around a BH surrounded by a spherically symmetric DM halo. 

%
\subsection{Background solution}
%
We consider a static, spherically symmetric spacetime 
\begin{equation}
ds^2 = g^{(0)}_{\mu\nu} dx^{\mu}dx^{\nu} = -a(r) dt^2 + \frac{dr^2}{1-\frac{2m(r)}{r}} + r^2 d\Omega^2\ ,
\end{equation}
where the metric functions $a(r),m(r)$ are solutions of 
the sourced Einstein field equations
\begin{equation}
    G_{\mu\nu}^{(0)} = 8\pi T^{m(0)}_{\mu\nu}\ .\label{eq:einsteineq}
\end{equation}
The stress-energy tensor $T^{m(0)}_{\mu\nu}$ is 
obtained following the Einstein cluster prescription~\cite{Einstein:1939ms} 
and describes the DM content as an anisotropic fluid:
\begin{equation}
    (T^{m(0)})^\mu_{\nu} = \text{diag}(-\rho, 0, P_t, P_t)\ .
\end{equation}
The density $\rho$ and the tangential pressure $P_t$ 
depend on the radial coordinate only, and are related through 
the Bianchi identities:
\begin{equation}
    P_t(r) = \frac{1}{2}\frac{m(r)}{r-2m(r)} \rho (r)\ .
\end{equation}
From the $tt$ and $rr$ components of Eqs.~\eqref{eq:einsteineq}, the mass function $m(r)$ and the metric function $a(r)$ must satisfy
\begin{equation}
    m'(r) = 4 \pi r^2 \rho(r)\,,\quad     \frac{a'(r)}{a(r)} = \frac{2 m(r)/r}{r - 2m(r)}\ .\label{eq:background}
\end{equation}

For a given density profile $\rho(r)$, solving Eqs.~\eqref{eq:background} fully specifies the background spacetime. We follow the numerical approach developed in Ref.~\cite{Figueiredo:2023gas} and consider three families of DM distribution \cite{Salucci_2019}, the Hernquist~\cite{1990ApJ...356..359H}, Navarro-Frenk-White (NFW)~\cite{Navarro:1996gj} and Einasto~\cite{1965TrAlm...5...87E,1969Afz.....5..137E} models.  The density profile of the first two can be described in terms of a two-parameter semianalytic function

\begin{equation}\label{eq:HQ_NFW_profiles}
\rho(r)=\rho_0(r/a_0)^{-\gamma}[1+(r/a_0)^\alpha]^{(\gamma-\beta)/\alpha}\ ,
\end{equation}
where $a_0$ is the scale radius of the halo, and the factor $\rho_0$
is proportional to the density, namely
$\rho_0=2^{(\beta-\gamma)/\alpha}\rho(a_0)$.  The parameters
$(\alpha,\beta,\gamma)$ are fixed to $(1,4,1)$ and $(1,3,1)$ for
Hernquist and NFW, respectively.  Given that the NFW distribution is
logarithmically divergent, we introduce a cutoff radius $r_c$, such
that $M_{\rm H}(r > r_c) = 0$, and assume $r_c= 5 a_0$.

The density profile for the Einasto model is given by
\begin{equation}
\rho(r)=\rho_e{\rm exp}\left\{-d_n[(r/r_e)^{1/n}-1]\right\}\ ,
\end{equation}
where $n=6$, $d_n=53/3$~\cite{Graham:2005xx,Prada:2005mx}, and
$\rho_e$ is the density at the radius $r_e$, that determines a volume
containing half of the halo mass.  Here we study the evolution of the
EMRI by setting $r_e=a_0$.

\begin{figure*}[!ht]
\centering
\includegraphics[width=1\linewidth]{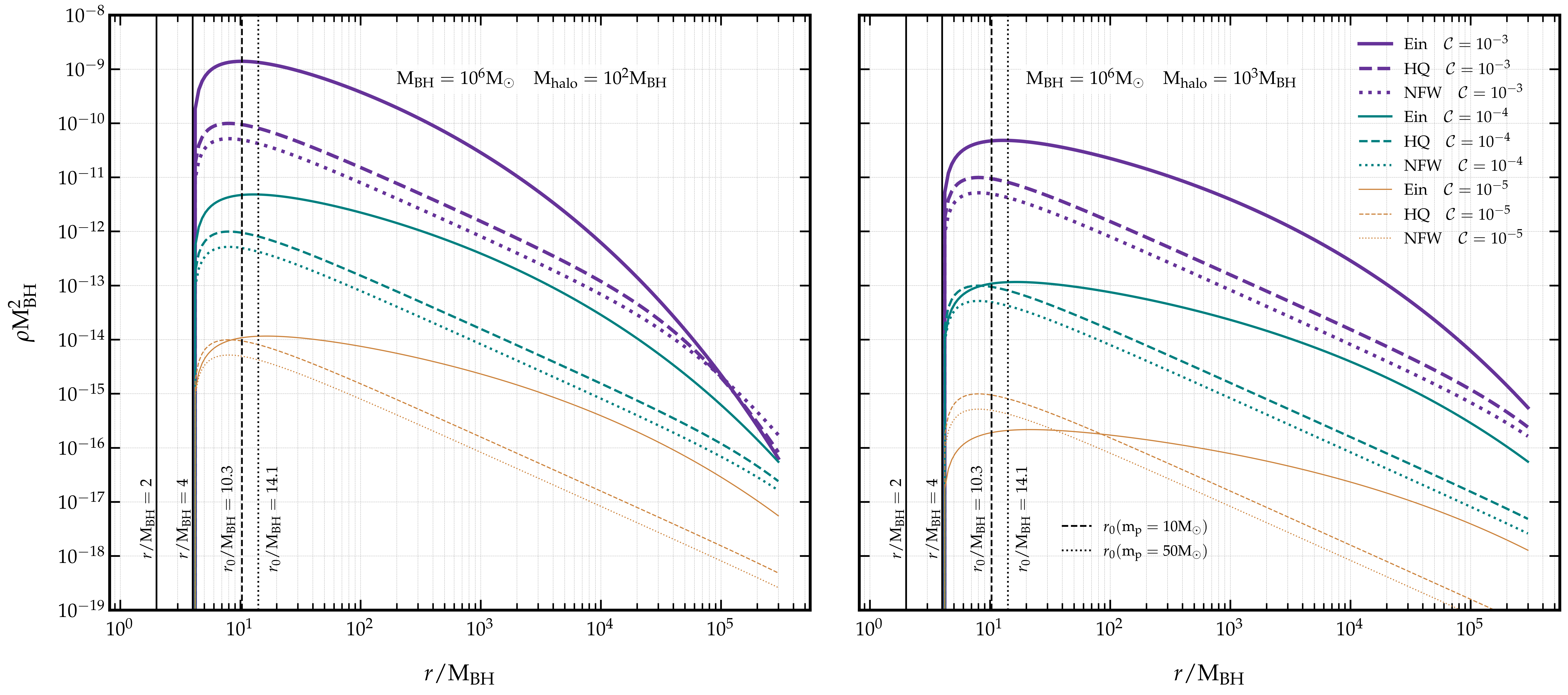}
\caption{Density profiles as a function of 
distance from a primary BH with mass $M = 10^6 M_\odot$. 
The purple-thick, green-medium, and orange-thin curves 
correspond to halo compactness values of $\mathcal{C} = 10^{-3}$, 
$\mathcal{C} = 10^{-4}$, and $\mathcal{C} = 10^{-5}$, 
respectively. The style of each curve indicates the DM model: Einasto (continuous), 
Hernquist (dashed), and NFW (dotted). The left and right panels correspond to DM halos 
with masses $M_\text{halo} = 10^{2} M_\text{BH}$ and 
$M_\text{halo} = 10^{3} M_\text{BH}$, respectively.  
Dashed and dotted vertical lines mark the initial orbital 
radius of a secondary object with mass $m_p = 10M_\odot$ 
and $m_p = 50M_\odot$, respectively, which evolves for 
one year around the primary before reaching the innermost stable circular orbit (ISCO). For reference, 
we also show in black the location of the coordinate 
radius of the BH horizon and the distance at which the 
DM profile vanishes.}
\label{fig:profiles}
\end{figure*}

Finally, for all profiles, we scale the density distribution by a
``peak factor'' such that $\rho(r)\rightarrow \rho(r)(1-2r_h/r)$.  This
factor is introduced to replicate the overdense cusps with a sharp
cutoff at $r=2r_h$, with $r_h=2M_{\rm BH}$ being the BH horizon, which
are expected to result from the adiabatic accretion growth of a
nonrotating BH~\cite{Sadeghian:2013laa,Gondolo:1999ef}. We require
the density to vanish for $r \le 2r_h$.

Realistic DM profiles, such as those inferred for Milky Way-like
galaxies, are expected to have small values of the compactness
${\cal C}\equiv M_{\rm H}/a_0\ll 1$.  In this regime of compactness,
as discussed in Refs.~\cite{Cardoso:2021wlq, Cardoso:2022whc},
deviations from the Schwarzschild vacuum solution scale linearly with
${\cal C}$. Unlike previous studies, in this work we consider low
density, realistic scenarios, focusing on DM halos with
${\cal C}=(10^{-3},10^{-4},10^{-5})$. Moreover, we study binary
systems evolving in DM halos with masses
$M_\text{halo} = (10^2, 10^3) M_\text{BH}$, where $M_\text{BH}$ is the
mass of the primary BH. Finally, to gauge the dependence of our
results on the NFW cutoff radius and on the value of $r_e$ for the
Einasto model, in Appendix~\ref{app:differentprofiles} we provide a
complementary analysis for some selected EMRI configurations, assuming
different values of $r_c$ and $r_e$.

In Fig.~\ref{fig:profiles} we show the density profiles of the nine different DM halo configurations analyzed in this work.
The scale and shape of the Hernquist and NFW profiles are similar. In
contrast, the Einasto model exhibits a steeper fall-off at large
distances from the BH, i.e., for $r \gtrsim 10^5 M_{\odot}$.  As a
result, the Einasto profile is more compact for a given $M_{\rm halo}$
and $\mathcal{C}$, with densities up to an order of magnitude higher
than the Hernquist and NFW distributions at distances
$r \sim 10^3 M_{\rm BH}$.

The dashed and dotted vertical lines in Fig.~\ref{fig:profiles}
indicate the initial orbital separation of EMRIs with a secondary mass
of $m_p = 10 M_\odot$ and $m_p = 50 M_\odot$, respectively, evolving
over one year until plunge (see Sec.~\ref{sec:setup}). This clearly
shows that typical EMRI configurations targeted by LISA observations
spend the majority of their orbital cycles near the peak of the DM
distribution.

%
\subsection{Gravitational and matter perturbations}
%
At the leading dissipative order, the EMRI dynamics is determined 
by linear perturbations sourced by the light BH 
orbiting around the massive primary:
\begin{equation}
    g_{\mu\nu} = g_{\mu\nu}^{(0)} + g_{\mu\nu}^{(1)} \quad \ ,\quad 
    T_{\mu\nu}^{m} = T_{\mu\nu}^{m(0)} + T_{\mu\nu}^{m(1)} \ ,
\end{equation}
which lead to the perturbed Einstein equations
\begin{equation}
    G_{\mu\nu}^{(1)} = 8\pi T_{\mu\nu}^{m(1)} + 8\pi T_{\mu\nu}^p\ ,
\end{equation}
where $ T_{\mu\nu}^p$ is the stress-energy tensor of the particle
(i.e., the secondary compact object).  The symmetry of the background
allows us to decompose $g^{(1)}_{\mu\nu}$, $T^{m(1)}_{\mu\nu}$ and
$T^{p}_{\mu\nu}$ into axial and polar modes, which can be treated
independently \cite{Regge:1957td,Davis:1971gg}.  We work in the
frequency domain, expanding perturbations in Fourier components, and
adopt the Regge-Wheeler-Zerilli gauge.

Matter and tensor modes couple in the polar sector only, with the DM
environment affecting axial perturbations only through background
quantities.  This allows us to obtain a single, second order,
differential equation for axial modes that differs from the
Schwarzschild case only in the values of $a(r)$ and $m(r)$. The polar
sector is more complex and cannot be reduced to a Zerilli-type
equation, requiring us to work with a set of coupled ODEs.

In the following we present a brief discussion of the relevant
formalism. We refer the reader
to~\cite{Cardoso:2022whc,Figueiredo:2023gas} for more technical
details.

%
\subsubsection{Axial perturbations}
%
The axial sector is fully described by two functions,
$h_0^{\ell m}(r,\omega)$ and $h_1^{\ell m}(r,\omega)$, where $\omega$
is the mode frequency, and the indices $(\ell,m)$ specify the
multipolar component of the perturbation. Einstein's equations
allow us to express $h_0^{\ell m}$ in terms of $h_1^{\ell
  m}$. Moreover, introducing the new variable
$\Psi_{\ell m}(r,\omega) = h_1^{\ell m}(r,\omega)
[a(r)(1-2m(r)/r)]^{1/2}/r$, we obtain a single, second order ODE:
\begin{equation}\label{eq:masteraxial}
    \frac{d^2 }{dr_*^2}\Psi_{\ell m} + [\omega^2 - V^{\text{ax}}_\ell(r)]\Psi_{\ell m} = S_{\ell m}^{\text{ax}}\ ,
\end{equation}
where the generalized tortoise coordinate $r_*$ is defined 
by
\begin{equation}
    \frac{dr_*}{dr} = a(r)^{-1/2}\left[1-\frac{2m(r)}{r}\right]^{-1/2}\ .
\end{equation}
The axial potential is
\begin{equation}
    V_\ell^{\text{ax}}(r) = \frac{a(r)}{r^2} \left[\ell(\ell+1) - \frac{6 m(r)}{r} + m'(r)\right]\ ,
\end{equation}
while the source term $S^{\text{ax}}_{\ell m}$ carries information
about the orbit of the secondary.

We solve Eq.~\eqref{eq:masteraxial} using Green's functions.  As a
first step we consider the associated homogeneous problem, with the
physical solution admitting purely ingoing and outgoing boundary
conditions at the horizon ($r=r_{\rm h}$) and infinity
$(r \rightarrow \infty)$, respectively. The solution at the boundaries
can be expanded as a power series:
\begin{equation}
    \Psi_{\ell m}^{(\rm in)} = e^{-i \omega r_*} \sum_{i = 0}^{n_{\text{in}}} \psi^{\rm in}_i (r - r_{\text{h}})^i\ ,
\end{equation}
\begin{equation}
    \Psi^{(\rm out)}_{\ell m} = e^{i \omega r_*} \sum_{i = 0}^{n_{\text{out}}} \psi^{\rm out}_i \frac{1}{r^i}\ ,
\end{equation}
in which we set $n_{\text{out}} = n_{\text{in}} = 4$. The values of  $(\psi^{\rm in}_i,\psi^{\rm out}_i)$ 
can be found by substituting $(\Psi_{\ell m}^{(\rm in)},\Psi_{\ell m}^{(\rm out)})$ into Eq.~\eqref{eq:masteraxial}, 
expanding in powers of $(r-r_h)$ and $1/r$ respectively, and solving order-by-order for the 
coefficients. The leading-order terms can be fixed to $\psi^{\rm in}_0=\psi^{\rm out}_0=1$. 

The full solution at infinity is obtained by integrating the homogeneous solution over the source term:   
\begin{equation}\label{eq:axialsol}
 \Psi^{\rm out}_{\ell m} = \Psi_{\ell m}(r_*\to\infty,\omega) = \frac{e^{i \omega r_*}}{W} \int_{r_\text{h}}^{r_\infty}\Psi_{\ell m}^{\rm (in)} S^{\rm ax}_{\ell m} dr_* \ ,
\end{equation}
where the Wronskian is defined as  
$W \equiv d\Psi_{\ell m}^{\rm out}/d r_* \Psi_{\ell m}^{\rm in} - d\Psi_{\ell m}^{\rm in}/d r_* \Psi_{\ell m}^{\rm out}$.
The axial energy flux at infinity is then given by
\begin{equation}
    \Dot{E}_{\ell m}^{\infty, \text{ax}}/q^2 = \frac{1}{8 \pi} \frac{(\ell+2)!}{(\ell-2)!} \vert \Psi^{\rm out}_{\ell m} \vert^2\ .
\end{equation} 

%
\subsubsection{Polar perturbations}
%
Polar modes are described by three metric functions,
$K^{\ell m} (r,\omega), H_0^{\ell m} (r,\omega)$ and
$H_1^{\ell m} (r,\omega)$, coupled with the fluid velocity
perturbation $W_{\ell m} (r,\omega)$ and the density component
$\delta\rho_{\ell m} (r,\omega)$.  A barotropic equation of state
provides a relation between pressure and density variations via the
radial and tangential speed of sound
\begin{equation}
c^2_{t,r}=\delta p_{t,r}^{\ell m}(r,\omega)/\delta\rho^{\ell m}(r,\omega)\ .
\end{equation}
Following~\cite{Cardoso:2022whc} we assume the speeds of sound to be
constant, although it generally depends on the radial coordinate $r$.
This leads to a set of five first-order ODEs, that can be cast in a
compact form as
\begin{equation}\
    \frac{d\Vec{\Phi}_{\ell m}}{dr} - \Hat{\alpha} \Vec{\Phi}_{\ell m} = \Vec{S}_{\ell m}\ ,\label{eq:polarEqs}
\end{equation}
where $\Vec{\Phi}_{\ell m} = \left\{K_{\ell m}, H_{0{\ell m}}, H_{1{\ell m}}, W_{\ell m}, \delta\rho_{\ell m} \right\}$, 
$ \Vec{S}_{\ell m} = (S_{1\ell m}, S_{2\ell m}, S_{3\ell m}, 0, 0)$, and the elements of the matrix $\Hat{\alpha}$ 
are shown in the Supplementary Material of Ref.~\cite{Cardoso:2022whc}.

The system \eqref{eq:polarEqs} is solved through a shooting method.
We first compute the appropriate boundary conditions of the full
inhomogeneous problem, which are purely ingoing at the horizon and
purely outgoing at infinity, respectively.  At the horizon, we adopt
the following ansatz:
\begin{align}
    K_{\ell m}^{(\rm in)} =& e^{-i\omega r_*} \sum_i^{n_{\rm in}} k^{\rm in}_i (r - r_\text{h})^i\ ,\\
    H_{0\ell m}^{(\rm in)} =& e^{-i\omega r_*} \sum_i^{n_{\rm in}} h^{\rm in}_{0_i} (r - r_\text{h})^{i-1}\ ,\\
    H_{1\ell m}^{(\rm in)} =& e^{-i\omega r_*} \sum_i^{n_{\rm in}} h^{\rm in}_{1_i} (r - r_\text{h})^{i-1}\ ,
\end{align}
while, following Ref.~\cite{Speeney:2024mas}, we fix
$W_{\ell m}(r_{\rm h}) = \delta\rho_{\ell m}(r_{\rm h}) =0$.
Similarly, at spatial infinity we introduce a Taylor expansion for the
metric functions
\begin{align}
    K_{\ell m}^{(\rm out)} =& e^{i\omega r_*} \sum_i^{n_{\rm out}} k^{\rm out}_i \frac{1}{r^{i}}\label{eq:Kout}\ ,\\
    H_{0{\ell m}}^{(\rm out)} =& e^{i\omega r_*} \sum_i^{n_{\rm out}} h^{\rm out}_{0_i} \frac{1}{r^{i-1}}\ ,\\
    H_{1{\ell m}}^{(\rm out)} =& e^{i\omega r_*} \sum_i^{n_{\rm out}} h^{\rm out}_{1_i} \frac{1}{r^{i-1}}\ ,
\end{align}   
and we set $ W_{\ell m}(r_\infty) = \delta\rho_{\ell m}(r_\infty) = 0$.
We fix $n_{\rm in} = n_{\rm out} = 5$. 
The coefficients at the horizon and at infinity can be computed with 
the same iterative procedure discussed for the axial sector. Setting 
$k^{\rm out}_0 = 1$ completely determines all the coefficients, modulo 
the shooting parameter $k^{\rm in}_0$.
For each radius $r$, the solution to Eqs.~\eqref{eq:polarEqs} is found 
by requiring the perturbations and their derivatives to 
be continuous at some  radius $r_\infty$, i.e.,
\begin{equation}\label{eq:polarsol}
    \lim_{r\to r_\infty} \left[K^{\ell m}_{\text{(sol)}} K^{'\ell m}_{\text{(out)}}-K'^{\ell m}_{\text{(sol)}} K^{\ell m}_{\text{(out)}}\right] = 0\ ,
\end{equation}
where a prime denotes differentiation with respect to $r$, 
$K_{\text{(sol)}}(r)$ is the numerical solution obtained 
by direct integration of the ODEs, and $K_{(\rm out)}$ is given by 
Eq.~\eqref{eq:Kout}.
Finally, the polar energy flux at infinity 
for each multipolar component is given by
\begin{equation}\label{eq:polarenergyflux}
    \Dot{E}_{\ell m}^{\infty, \text{pol}}/q^2=\lim_{r\to r_{\text{obs}}} \frac{1}{32 \pi} \frac{(\ell+2)!}{(\ell-2)!}|K_{\ell m}(r)|^2\ .
\end{equation}
%

%
\subsection{Geodesics and orbital evolution}
%
Given the background functions $a(r)$ and $m(r)$, we can determine the
geodesic properties of particles in the DM halo. The spacetime admits
two conserved quantities, corresponding to the energy per unit mass
and the specific angular momentum at infinity:
\begin{equation}
\label{eq:enr_ang}
E_p=\left[ \frac{r-2m(r)}{r-3m(r)}a(r) \right]^{1/2}_{r=r_p}\ , ~~~ L_p=\left[ \frac{m(r)}{r-3m(r)} \right]^{1/2}_{r=r_p}\ ,
\end{equation}
where $r_p$ denotes the orbital radius of the particle, and its
orbital frequency is given by $\omega_p= a(r_p) L_p/r_p^2 E_p$.  We
focus on equatorial orbits, and set $\theta(r=r_p)=\pi/2$. From
$m(r)$ we can also compute the radius of the last stable orbit for
massive bodies, $r_{\rm isco}$, which satisfies the following
equation:
\begin{equation}\label{eq:ISCOmat}
r^2m'(r)+rm(r)-6m(r)^2=0\ .   
\end{equation}

Once we have the solution for the background and the GW energy fluxes,
we can numerically integrate the equations of motion for the orbital
phase and separation:
\begin{equation}\label{eq:eom1}
       \frac{d\Phi}{dt} = \omega_p\quad\ ,\quad 
 \frac{dr}{dt} = - \Dot{E}\frac{dr}{dE_\text{orb}}\ ,
\end{equation}
where $E_\text{orb}$ is the orbital energy of the particle.

We solve Eqs.~\eqref{eq:eom1} with initial conditions $\Phi(t=0)=0$
and $r(t=0)=r_0$, where $r_0$ is chosen such that the EMRI reaches the
plunge radius $r_{\rm plunge}$ after time period $T$. We make the
conservative choice $r_{\rm plunge}=r_{\rm isco}+0.1M_\text{BH}$
because the precision in computing the fluxes decreases for orbits
close to $r=r_{\rm isco}$. We will present results for two cases:
$T=1$~year and $T=6$~months.

%
\section{Numerical setup and 
detectability}\label{sec:setup}
%

The total energy flux emitted by the binary at each radius is given by
\begin{equation}
    \dot{E} = \dot{E}^{\infty} + \dot{E}^{H},
\end{equation}
where \(\dot{E}^{\infty}\) and \(\dot{E}^{H}\) denote 
the energy fluxes at infinity and at the horizon, respectively. Here we will neglect the contribution from the horizon fluxes. First of all, including them 
in our setup would require a different numerical 
scheme, as their accurate extraction through the 
shooting method for polar modes becomes challenging 
near \(r \to r_h\). Secondly, we have explicitly verified, 
using direct integration methods for the axial sector, 
that including the horizon fluxes produces relative dephasings between matter and vacuum 
evolution smaller than $0.01\%$.
Given that the ratio of axial to polar 
energy fluxes at infinity is similar, 
we expect the contribution from the polar 
horizon fluxes to be similarly small, and we conclude that 
horizon fluxes should be negligible for the configurations we consider.
We also note that no matter fluxes are present in our model. The matter distribution is modeled as a stationary, gravitationally bound, and non-dissipative system. Although its internal energy may vary through coupling with gravitational perturbations, this energy is merely redistributed within the fluid. The only genuinely radiative degrees of freedom are those of the gravitational field itself. A net energy flux associated with matter would require dissipative processes such as viscosity or mass shedding, which are not included in our stress-energy tensor.
The interaction between the secondary and the fluid may in principle generate local effects, 
such as drag forces, whose consistent relativistic treatment would require a self-force approach 
and lies beyond the scope of this work. Hence, we explicitly assume that the binary evolution 
is driven by a balance law that includes only GW fluxes.

The total flux at infinity is then given by
\begin{equation}
    \Dot{E}^{\infty} = \sum_{\ell = 2}^{\ell_{\text{max}}}\sum_{m=-\ell }^{\ell} (\Dot{E}_{\ell m}^{\infty, \text{ax}} + \Dot{E}_{\ell m}^{\infty, \text{pol}})\ .\label{eq:energyfluxes}
\end{equation}
The results presented in the main text have been obtained by setting
$\ell_\text{max} = 2$. In Appendix~\ref{app:lmax}, we validate this
choice by comparing the results for different values of
$\ell_\text{max}$. The values of 
$\Dot{E}_{\ell m}^{\infty}$ are
computed on a uniform grid of points in the range
$r\in[6.0001,14.5001]M_{\text{BH}}$, with step size of
$r_\text{step} = 0.1 M_{\text{BH}}$.  We solve the equations of motion
\eqref{eq:eom1} on this grid for a primary mass $M=10^6 M_\odot$ and
two values of the secondary mass: $m_p=(10,50)M_\odot$.  We extract
fluxes by setting the numerical infinity in Eqs.~\eqref{eq:axialsol}
and \eqref{eq:polarsol} to $r_\infty = 2 \times 10^7 M_{\rm{BH}}$ when
$M_{\rm{halo}} = 10^2 M_{\rm{BH}}$, and
$r_\infty = 2 \times 10^8 M_{\rm{BH}}$ when
$M_{\rm{halo}} = 10^3 M_{\rm{BH}}$. These values correspond to
$2 a_{0_{\rm{max}}}$, where $a_{0_{\rm{max}}}$ is the maximum value of
$a_0$ among the three different choices used to define the compactness
for a fixed value of $M_{\rm{halo}}$ (we have also checked that
changes in $r_\infty$ by one order of magnitude do not significantly
affect the GW fluxes).  Finally, to ensure a proper comparison with
vacuum fluxes (and consequently with GW phases and waveforms), we
compute the vacuum fluxes at the same numerical infinity.

\begin{figure*}[t]
\centering
\includegraphics[width=1\textwidth]{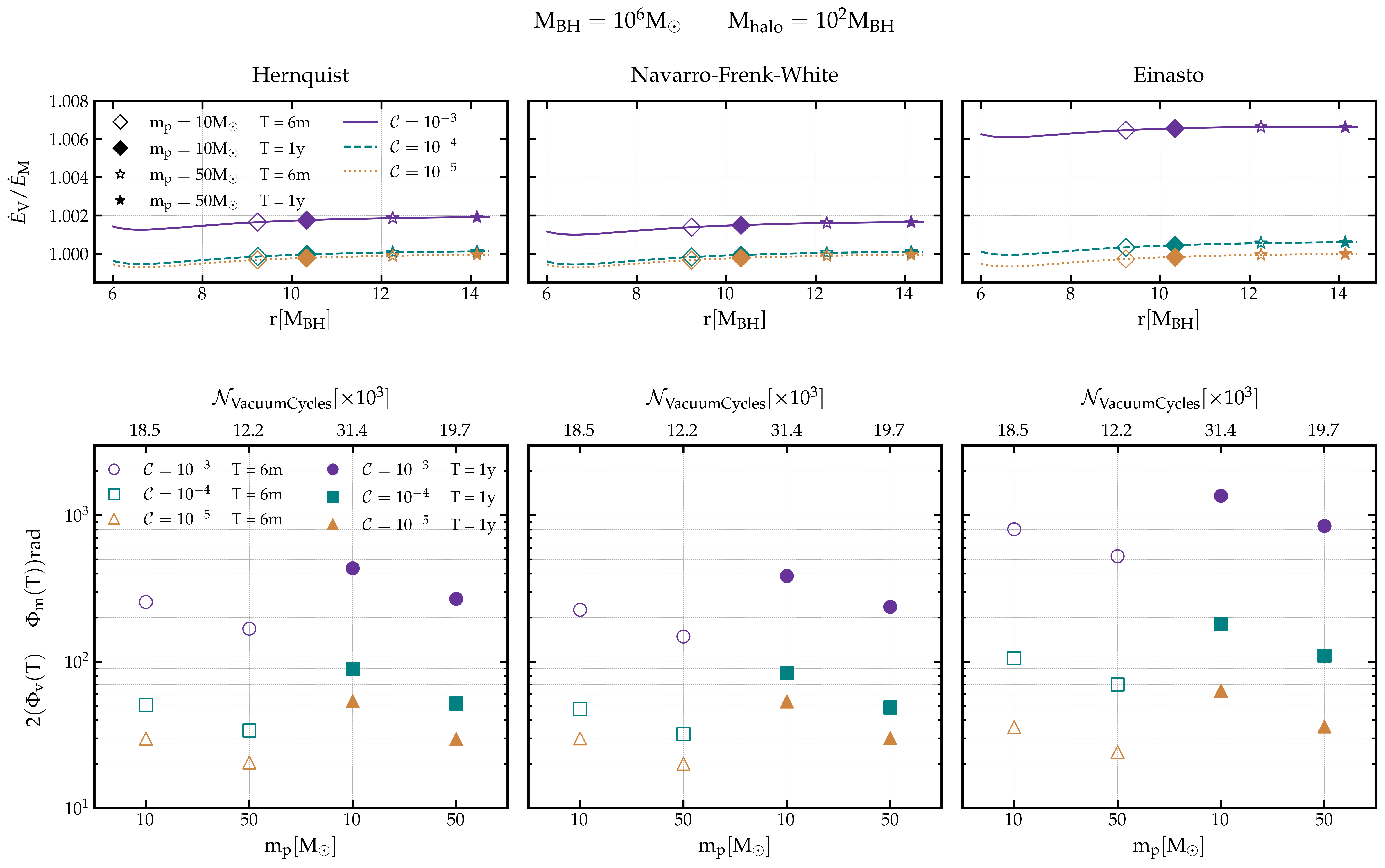}
\caption{Top row: Ratio of GW fluxes emitted in vacuum to those in the
  presence of matter, plotted as a function of the secondary’s orbital
  radius. The primary BH mass is \( M = 10^6 M_\odot \), the DM halo
  mass is \( M_{\text{halo}} = 10^{2} M_{\text{BH}} \), and we
  consider two possible values of the secondary mass:
  \( m_P = (10,50) M_\odot \).  The purple solid, green dashed, and
  orange dotted lines represent halos with compactness values of
  \( \mathcal{C} =(10^{-3},10^{-4},10^{-5}) \).  Markers along each
  curve indicate the initial orbital radius of the secondary, which
  evolves over either one year (filled markers) or six months (hollow
  markers) until plunge (see legend). Bottom row: Quadrupolar
  dephasing accumulated over one year (filled markers) and six months
  (hollow markers) during the evolution of the EMRIs. Labels on the
  top axes indicate the accumulated number of cycles in the vacuum
  case. The three different panels, from left to right, correspond to
  different DM density profiles.}
\label{fig:dephasing_Mh100}
\end{figure*}

\begin{figure*}[t]
\centering
\includegraphics[width=1\textwidth]{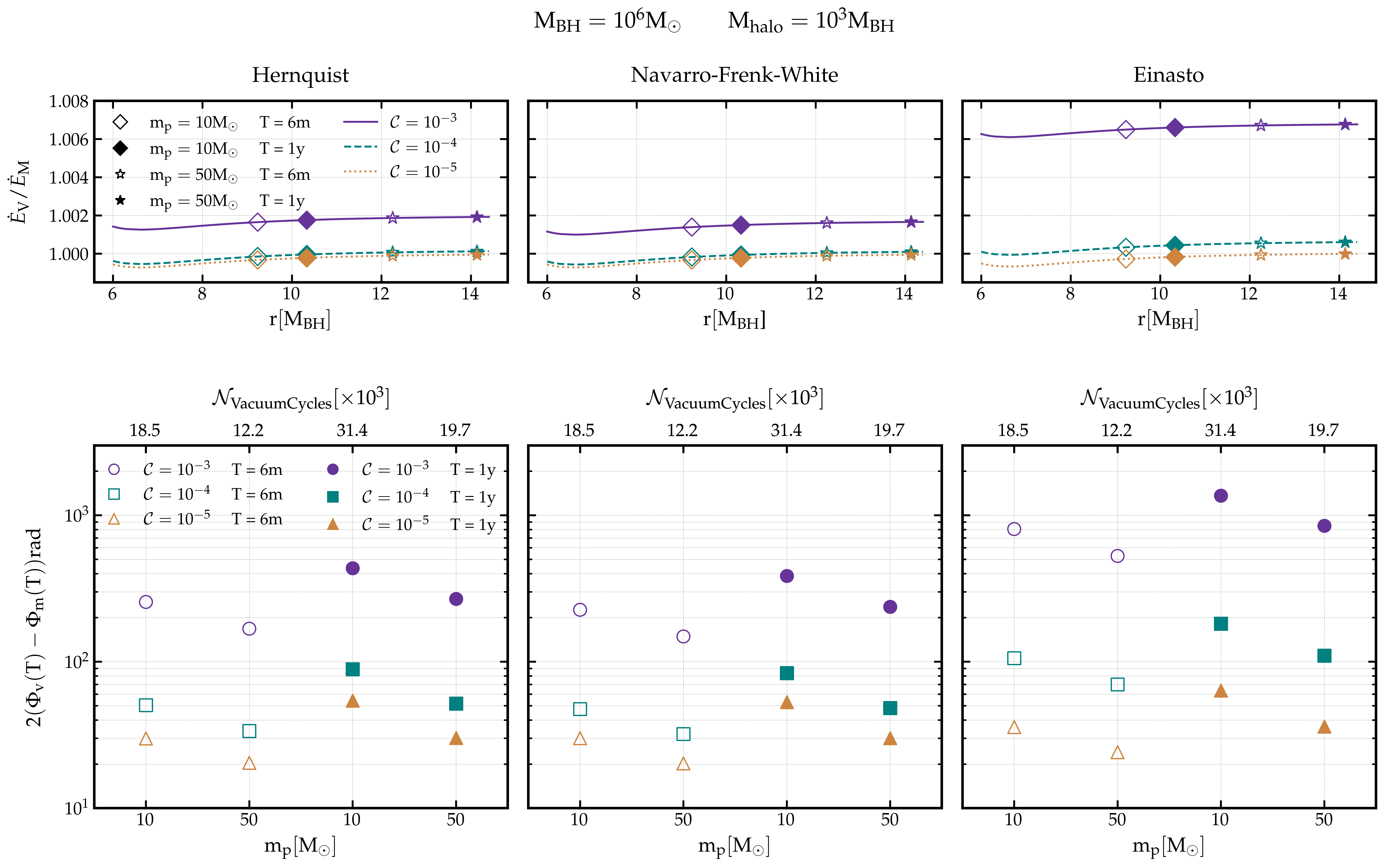}
\caption{Same as Fig.~\ref{fig:dephasing_Mh100}, but assuming a DM
  halo with $\text{M}_\text{halo} = 10^{3} \text{M}_\text{BH}$.}
\label{fig:dephasing_Mh1000}
\end{figure*}

To assess the environment detectability we consider two figures of
merit. We first study the evolution in phase of the GW signal,
computing the phase difference between binaries in vacuum and in a DM
background, $\Delta\Phi=\Phi_v-\Phi_m$, after a given observation time
$T$. A value of $\Delta \Phi$ larger than a certain threshold provides
an indication that the environment can be detected by LISA. Following
\cite{Bonga:2019ycj}, we assume the threshold to be
$\Delta\bar{\Phi}=0.1$ radians, which roughly corresponds to a
resolvable phase for a system with signal-to-noise ratio of $\sim 30$.

A more quantitative and accurate assessment can be made by evaluating
waveform changes weighted by the detector sensitivity. To this aim we
compute the faithfulness ${\cal F}$ between GW signals from
EMRIs in vacuum, and those surrounded by the DM distribution:

\begin{equation}\label{eq:faith}
    \mathcal{F}\left[h_m,h_v\right] = \max_{\{t_c,\phi_c\}} \frac{{\langle h_m|h_v\rangle}}{\sqrt{{\langle h_m|h_m\rangle}{\langle h_v|h_v\rangle}}}\ ,
\end{equation}
where 
\begin{equation}\label{eq:scalar}
    {\langle h_m|h_v\rangle} = 4 \Re \left[ \int_{f_{\text{min}}}^{f_{\text{max}}} \frac{\Tilde{h}_m(f)\Tilde{h}_v^*(f)}{S_n(f)} df\right]\,.
\end{equation}

Here $S_n(f)$ is the LISA noise power spectral density, including the
confusion noise produced by unresolved galactic white dwarf
binaries~\cite{Robson:2018ifk}. The frequency domain waveforms
$\tilde{h}_{m,v}(f)$ are computed by applying a discrete Fourier
transform to the time domain plus and cross signal polarizations.
These are derived using the quadrupolar
formula~\cite{Huerta:2011kt,Barack:2003fp}, i.e.,
\begin{equation}
h_+=(\ddot{I}_{11}-\ddot{I}_{22})/2 \ ,\quad 
h_\times=\ddot{I}_{12}\ ,
\end{equation}
where a dot represents a derivative with respect to the coordinate
time, the quadrupole moment tensor is defined as
$I_{ij}=2m_p/d_{\rm L} z^i(t)z^j(t)$, $z(t)$ is the geodesic of the
secondary expressed in Cartesian spatial coordinates, and $d_{\rm L}$
is the source luminosity distance.  We refer the reader to the method
section of~\cite{Maselli:2021men} for further technical details on the
waveform generation, and on the numerical implementation of
Eq.~\eqref{eq:faith}. Finally, we fix $f_{\text{min}} = 10^{-4}$\,Hz
in the integral of Eq.~\eqref{eq:scalar}, while $f_\text{max}$
corresponds to the Nyquist frequency of the signal.  We assume that
two waveform models are distinguishable if the faithfulness is smaller
than the critical value $\bar{{\cal F}}\simeq0.95$, corresponding to
the estimated distinguishability threshold for a template depending on
ten parameters, observed with a signal-to-noise ratio of
$\sim10$~\cite{Chatziioannou:2017tdw}.

%
\section{Results}\label{sec:results}
%

We assess the detectability of three families of 
DM halos: Hernquist, NFW, and Einasto models. We assume halo 
compactness values of $\mathcal{C} = (10^{-3}, 10^{-4}, 10^{-5})$, 
which are lower than those in previous studies and provide a more realistic 
description of DM densities within overdensities extending up to $10^7$ 
gravitational radii from the central BH.

In Fig.~\ref{fig:dephasing_Mh100} (top row) we show the ratio of
matter to vacuum fluxes for EMRIs evolving in DM halos with
$M_{\rm halo} = 10^2 M_{\text{BH}}$ 
as a function of the secondary’s
orbital radius. Filled (hollow) markers represent systems evolving for
one year (six months) in the LISA band before plunge, respectively,
with each marker positioned at the initial inspiral radius.
Deviations from the Schwarzschild case can reach up to $10^{-3}$, and
generally increase (if only mildly) as a function of the initial
orbital separation.

In many cases, fluxes in matter are lower than those in vacuum for
orbital separations close to the ISCO.  As shown in
Refs.~\cite{Cardoso:2022whc}, this reduction can be partially
explained in terms of the gravitational redshift induced at the
leading order in the halo compactness by the extended distribution of
DM. However, $\dot{E}_{v} / \dot{E}_{m}$ has a local minimum at
$r \sim 7 M_{\rm BH}$ for secondaries closer to the BH. We attribute
this to the DM radial distribution, which peaks just below
$r \sim 10 M_{\rm BH}$ (see Fig.~\ref{fig:profiles}). In regions of
higher DM density, nonlinear contributions from $M_{\rm halo}$ and
$a_0$ may become more relevant, counteracting redshift effects.  The
ratio $\dot{E}_{v} / \dot{E}_{m}$ tend to grow again as $r$ approaches
$r_{\rm ISCO}$, where the density $\rho(r)$
decreases.

While the redshift scaling is the primary effect for axial fluxes, the
fluxes in the polar sector can also be affected by couplings between
the metric and fluid modes. However, we have verified that for the
values of ${\cal C}$ considered here such couplings have a very small
effect. We have numerically integrated the ODE
system~\eqref{eq:polarEqs} by artificially setting matter
perturbations to zero, and we find values of $\dot{E}_{m}$ almost
indistinguishable from those in Fig.~\ref{fig:dephasing_Mh100}.  For
small values of the halo compactness, changes in the EMRI evolution
are primarily driven by the non-Schwarzschild background in which the
secondary moves.

The flux differences accumulate throughout the inspiral due to the
large number of orbits and they affect the binary's phase evolution,
as shown in the bottom panels of Fig.~\ref{fig:dephasing_Mh100}. For
all the configurations analyzed here the dephasing lies well above the
detection threshold, reaching values as high as $\Delta\Phi \sim 10^3$
radians over one year. As expected, the dephasing increases with
${\cal C}$, i.e., for more compact DM distributions.  We find that a
linear fit $\Delta\Phi = a + b {\cal C}$ describe all our data with an
accuracy better than $1\%$, regardless of the halo model and binary
configuration.
EMRIs with lighter secondaries accumulate a larger number of cycles
(as shown on the top x-axis of each panel), and therefore they have
the largest dephasings $\Delta\Phi$.

Comparing different halo models, Hernquist and NFW yield similar
results, while the Einasto distribution leads to the largest values of
$\Delta\Phi$. This is because the axial and polar fluxes are
affected by the local value of $\rho(r)$ appearing in
Eqs.~\eqref{eq:masteraxial} and \eqref{eq:polarEqs}. The radial
profiles in Fig.~\ref{fig:profiles} show that the densities of
Hernquist and the NFW models are very close -- within a factor two in
the region in which the secondary inspirals. In contrast, the Einasto
distribution has higher densities, exceeding Hernquist and NFW by a
factor $\sim 10-30$ within a distance of $\sim 10M_{\rm BH}$ from the
ISCO.

\begin{figure*}[!ht]
\centering
\includegraphics[width=1\textwidth]{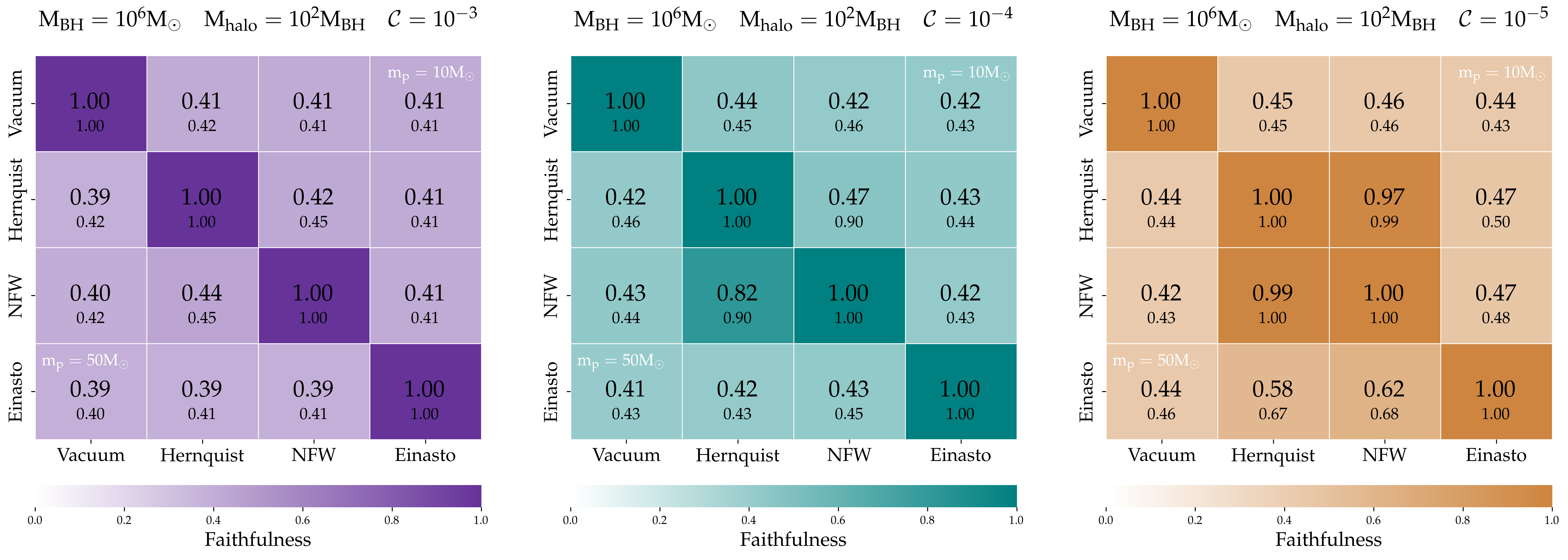}
\caption{Faithfulness density matrices for different DM profiles
  (Hernquist, NFW, and Einasto) of mass
  \( M_{\text{halo}} = 10^{2} M_{\text{BH}} \), compared to the vacuum
  case and among each other.  The three panels correspond to halo
  compactness values of
  \( \mathcal{C} =(10^{-3}, 10^{-4}, 10^{-5}) \). The upper and lower
  diagonals of each matrix display faithfulness values computed for
  systems with \( (M_{\text{BH}}, m_p) = (10^6,10) M_\odot \) and
  \( (M_{\text{BH}}, m_p) = (10^6,50) M_\odot \), respectively.  Large
  (small) numbers in each cell indicate faithfulness values computed
  for a one-year (six-month) evolution, respectively.}
\label{fig:faithfulness_Mh100}
\end{figure*}

\begin{figure*}[!ht]
\centering
\includegraphics[width=1\textwidth]{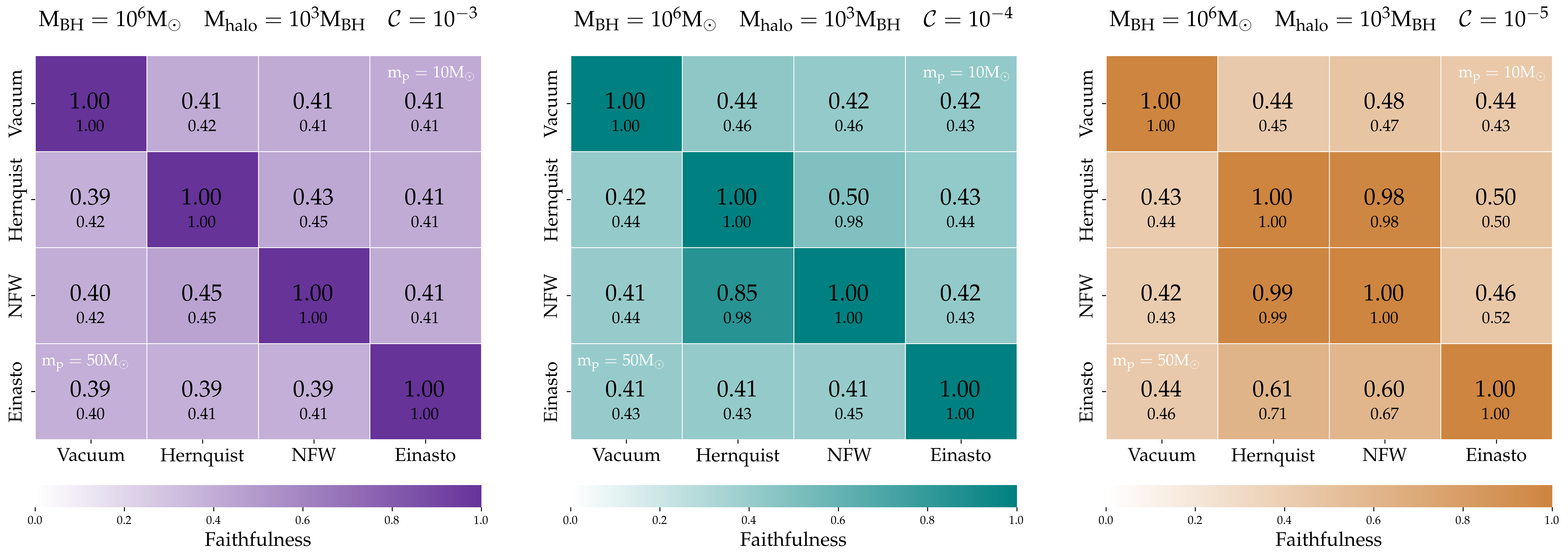}
\caption{Same as Fig.~\ref{fig:faithfulness_Mh100}, but for a halo
  mass of $M_\text{halo} = 10^{3} M_\text{BH}$.}
\label{fig:faithfulness_Mh1000}
\end{figure*}

In Fig.~\ref{fig:dephasing_Mh1000} we repeat the dephasing analysis
for halos of 
$M_{\rm halo}=10^3M_\text{BH}$ . 
The results are qualitatively
and quantitatively similar to those for lighter DM distributions.
This confirms that the phase evolution in the halo is primarily driven
by its compactness, rather than the total halo mass and size.  

To further support this conclusion, we have computed the GW fluxes and
evolved EMRIs assuming halos with the same compactness as the models
discussed so far. However, we considered different combinations of $M$
and $a_0$, specifically by setting $M_H = M_{\rm BH}$ and
$a_0 = (10^3,10^4,10^5)M_H$.  When comparing the dephasing relative to
vacuum evolutions, we found minor differences (of the order of a few
percent) only for the largest valus of the compactness,
${\cal C} = 10^{-3}$. Halos with lower compactness did not lead to any
significant changes compared to the EMRIs previously analyzed.

To assess the impact of the redshift correction, 
we computed redshifted frequencies\footnote{Since 
the binary orbital frequency is related to the radius 
by $\Omega = r^{-3/2}$  the radius also transforms as 
$r_r = \frac{r_v}{(1 + \delta z)^{2/3}}$.} 
and vacuum fluxes following the prescription of Ref.~\cite{Cardoso:2022whc}, such that 
$\dot{E}^\infty_r \rightarrow (1 - \delta \mathcal{C})\,\dot{E}_v$, and
$\Omega_r \rightarrow \Omega_v (1 + \delta z)$.
The numerical factor \(\delta\) depends on the DM 
density profile; we adopted values of 
$\delta = 1,0.9,3$ for the Hernquist, NFW, and Einasto 
profiles, respectively, following the results of 
Ref.~\cite{Figueiredo:2023gas} for the axial sector. 
For polar perturbations, obtaining analytical 
estimates of $\delta$ is more challenging. However, 
we have verified that varying $\delta$ around 
these fiducial values does not  
qualitatively change our results. 

As expected, the dephasing decreases when the redshift correction is included (see Table \ref{tab:redishift_vacuum} ), but it remains significant. This indicates that, while couplings with matter in the emitted fluxes are local, their secular effects influence the long-term evolution of the binary. 

\begin{table*}
\begin{tabular}{p{0.06\textwidth} p{0.06\textwidth} p{0.06\textwidth} p{0.16\textwidth} p{0.16\textwidth} p{0.16\textwidth}}
 &  &  & $C = 10^{-3}$  & $C = 10^{-4}$  & $C = 10^{-5}$  \\
 \hline
\hline
{\small density} &  $m_p$ & {\small T} & $\Delta \Phi=$  & $\Delta \Phi=$  &$\Delta \Phi=$  \\
{\small profile} &  $[M_\odot]$ & [year]  & $\Phi_m-\Phi_v(\Phi_r)$& $\Phi_m-\Phi_v(\Phi_r)$ &  $\Phi_m-\Phi_v(\Phi_r)$ \\
\hline
\hline
{\small Hernquist} & 10 & 1 & 436 (299)&89 (75)&54 (52) \\
                  &    & 0.5 & 256 (189)&50 (44)&30 (29)  \\
                  & 50 & 1 & 269 (149)&52 (40)&30 (29) \\
                  &    & 0.5 & 168 (103)&34 (27)&21 (20) \\
\hline
{\small NFW} & 10 & 1 & 386 (263)&84 (72)&54 (52) \\
             &    & 0.5 & 226 (130)&48 (42)&30 (29)  \\
             & 50 & 1 & 237 (166)&49 (38)&30 (29)  \\
             &    & 0.5 & 149 (90)&32 (26)&20 (20) \\
\hline
{\small Einasto} & 10 & 1 & 1360 (953)&182 (141)  &64 (60)  \\
                 &    & 0.5 & 803 (604)&106 (74)  &36 (34) \\
                 & 50 & 1 & 846 (489)&110 (86)  &36 (33)\\
                 &    & 0.5 & 526 (331)&70 (50)  &24 (22) \\
\hline
\hline
\end{tabular}
\caption{GW dephasing $\Delta\Phi$ with respect to vacuum (redshifted vacuum) computed for EMRIs with a primary of $M=10^6M_\odot$ and different secondary masses
  ($m_p=10 M_\odot$ or $m_p=50M_\odot$), evolving for either one year or six months in different DM profiles. We assume halo masses of  $\rm{M}_{\rm{halo}} = 10^2 \rm{M}_\odot$.}
  \label{tab:redishift_vacuum}
\end{table*}

While promising, these results do not account for parameter
correlations in the waveform model or the impact of LISA noise. To
assess the relevance of halo effects in realistic GW observations, we
have computed the faithfulness ${\cal F}$ defined in
Eq.~\eqref{eq:faith} between 0.95 in matter and vacuum for the
EMRI configurations analyzed above. Our results are summarized in the
matrix diagrams of Figs.~\ref{fig:faithfulness_Mh100}
and~\ref{fig:faithfulness_Mh1000}, where we list ${\cal F}$ for six
months and one year of observation,
$M_{\rm halo} = (10^2,10^3) M_{\odot}$, and for the three compactness
values considered above. For both $m_p=10M_\odot$ and $m_p=50M_\odot$,
GW signals from binaries evolving in DM environments differ
significantly from vacuum, with ${\cal F}\ll \bar{{\cal F}}$ (where
$\bar{{\cal F}}=0.95) $ is the distinguishability threshold discussed in
Sec.~\ref{sec:setup}) even in the case of diluted distributions with
${\cal C}=10^{-5}$.

The off-diagonal blocks in the matrices of
Figs.~\ref{fig:faithfulness_Mh100} and~\ref{fig:faithfulness_Mh1000}
show the faithfulness computed by comparing different halo profiles
against each other. They suggest that LISA observations may not only
distinguish matter effects from vacuum, but also discriminate between
different radial density profiles. Even in the case of signals with
similar halo models, like Hernquist and NFW, we obtain ${\cal F}<1$ in
most cases, with the exception of halos with ${\cal C}= 10^{-5}$.

The faithfulness is typically mildly affected as we vary the secondary
mass from $m_P=10M_\odot$ to $m_P=50M_\odot$. In general we would
expect the faithfulness for one-year observations to be smaller than
the for six-month observations, but this is not always the case. This
may be due to subtle variations in the waveform phase between the
different matter distributions, which can sometimes lead to larger
overlaps when the observation time is longer.

By comparing Figs.~\ref{fig:faithfulness_Mh100}
and~\ref{fig:faithfulness_Mh1000}, we see that these broad conclusions are insensitive to variations in the halo mass. 

Finally, we verified that all results obtained from the faithfulness analysis remain unchanged when using redshifted fluxes, as done for the dephasing. This confirms that the distinguishability of the signals persists even when the redshift effect is accounted for.

\begin{figure}[t]
\centering
\includegraphics[width=1\linewidth]{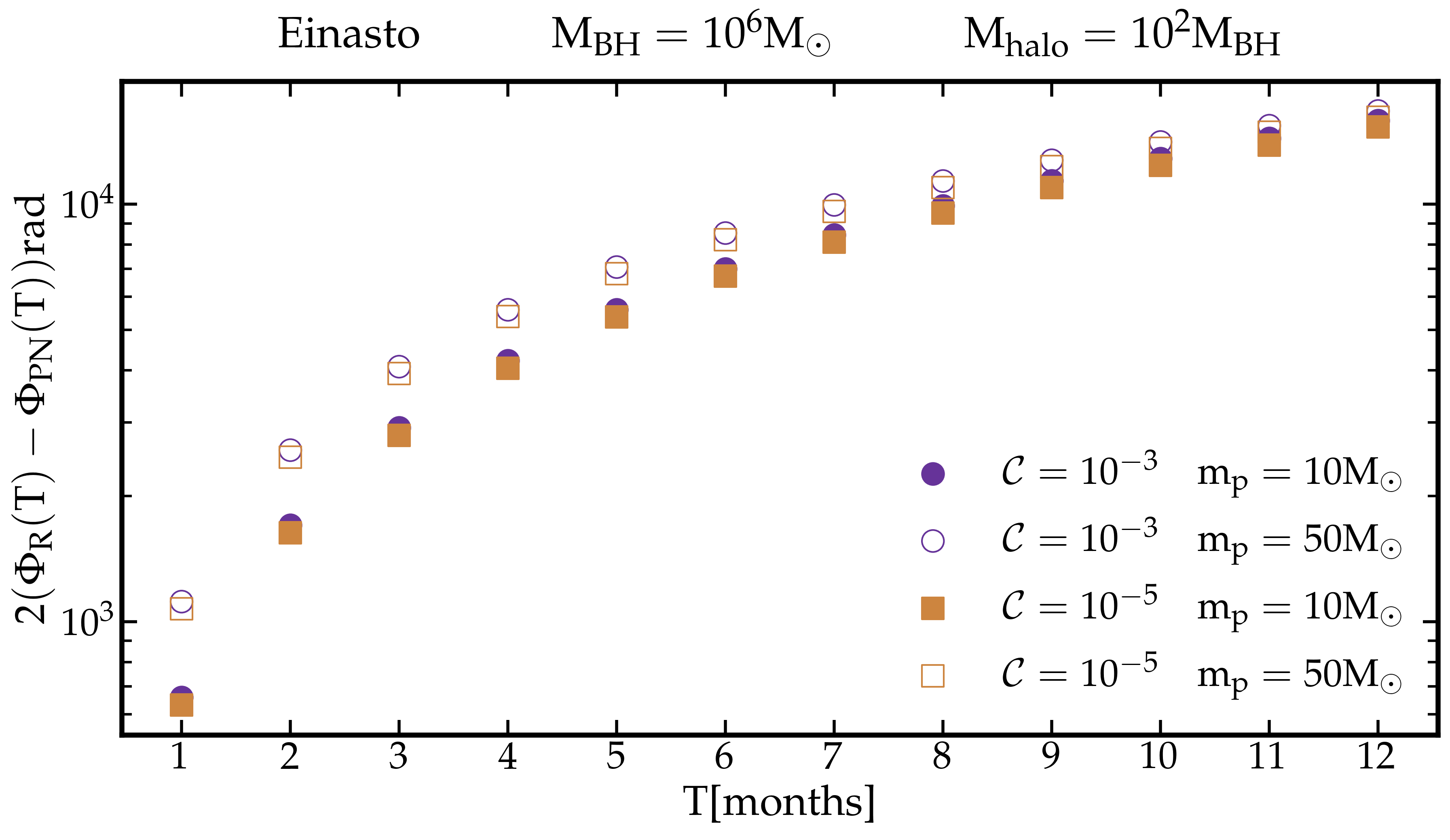}
\caption{Quadrupolar dephasing accumulated over one year of evolution
  for EMRIs with a primary mass of \( M = 10^6 M_\odot \) and a
  secondary of \( m_p = 10 M_\odot \) (filled markers) or
  \( m_p = 50 M_\odot \) (hollow markers). The binaries evolve within
  a dark matter halo described by the Einasto profile, with a halo
  mass of \( M_{\text{halo}} = 10^{2} M_{\text{BH}} \) and compactness
  values of \( \mathcal{C} = 10^{-3} \) (purple round markers) and
  \( \mathcal{C} = 10^{-5} \) (orange square markers).  The dephasing
  is computed by comparing EMRIs evolving under two different
  prescriptions for the GW energy flux. In one case, we use the
  \( l=2, m=2 \) component of the relativistic flux from
  Eq.~\eqref{eq:polarenergyflux}. In the other, we apply the
  quadrupolar PN expression given by Eq.~\eqref{eq:energyPN}.}
\label{fig:dephasing_PN}
\end{figure}

As a final remark, we want to emphasize the importance 
of using a fully relativistic description to model EMRI 
dynamics within DM halos. To demonstrate this, we 
have computed the phase difference between binaries 
in which the orbital elements in Eqs.~\eqref{eq:eom1}
evolve with two different schemes: one, $\Phi_{\text{R}}$,   based on the \( \ell = m = 2 \)
component of the GW flux in Eq.~\eqref{eq:energyfluxes}, and the other,  $\Phi_{\text{PN}}$,
using the quadrupolar post-Newtonian (PN) prescription
from Ref.~\cite{Speeney:2022ryg}:

\begin{equation}\label{eq:energyPN}  
\dot{E}_{\rm GW}^\infty = \frac{32}{5}(M_{\rm BH}\omega_p)^{10/3} \left[1+\frac{4}{3}\delta(r)\right] \ ,  
\end{equation}  
where \( \delta(r) = M_{\rm halo}(r) / M_{\rm BH} \), with
\( M_{\rm halo}(r) \) representing the mass of the halo enclosed
within a sphere of radius \( r \). In Fig.~\ref{fig:dephasing_PN} we
show the accumulated dephasing as a function of time until plunge, for
a one-year observation of EMRIs with a primary mass of
\( M = 10^6 M_\odot \) and a secondary of \( m_p = (10,50) M_\odot
\). We consider a DM halo modeled by the Einasto profile
with two different values of $\mathcal{C}$, but we have checked that
other DM profiles yield similar results. Even when
$\mathcal{C} = 10^{-5}$ the PN approximation quickly becomes
inadequate, with the accumulated phase difference \( \Delta\Phi \)
reaching hundreds of radians within just a few months of evolution.

%
\section{Conclusions}
%

In this work, we have studied the relativistic dynamics of EMRIs
evolving within DM distributions. We computed the gravitational and
fluid perturbations induced by a light compact object orbiting a
nonrotating BH embedded in a spherically symmetric halo, evaluating
the GW energy flux and the binary phase evolution for different halo
models.  We have analyzed three widely used DM models (the Hernquist,
Navarro-Frenk-White, and Einasto distributions) with different total
mass and characteristic scale and different binary masses, assessing
the ability of LISA observations to detect deviations from vacuum
predictions.

We have found that EMRIs evolving in DM environments can accumulate
phase differences of hundreds of radians compared to their vacuum
counterparts over a one-year evolution in the LISA band. For a given
DM profile, this dephasing is primarily determined by the halo
compactness, scaling linearly with ${\cal C}$. Lighter secondaries
experience larger phase deviations due to their slower inspiral and
higher number of accumulated cycles. For a fixed compactness, the
phase difference $\Delta \Phi$ depends on the maximum local density in
the orbital region where the secondary evolves. Even in diluted
configurations with ${\cal C} = 10^{-5}$, EMRIs with a secondary mass
of either $10M_\odot$ or $50M_\odot$ would experience a phase shift of
approximately 10 radians after roughly two months of observation.

These findings are further supported by a refined analysis we
performed to assess the distinguishability of GW signals weighted by
the LISA sensitivity curve, in which we computed the faithfulness
between waveforms emitted in vacuum and in the presence of DM. We find
that six months of observation are already sufficient to confidently
discriminate between EMRIs in vacuum and EMRIs evolving in DM
profiles. More interestingly, we find that LISA can also differentiate
among DM profiles with different radial density distributions.
Furthermore, by including the redshift correction, we have verified that the associated changes in the orbital frequency and energy flux tend to reduce the dephasing; however, they do not alter the qualitative behavior of the signal. This confirms that the cumulative impact of local matter couplings over time remains significant for the long-term orbital evolution.

A complete assessment of the ability of future observations to
constrain DM models, however, requires dedicated parameter estimation
techniques. The waveform models computed in this work can be readily
incorporated into LISA data analysis pipelines such as Fast EMRI
Waveforms~\cite{Katz:2021yft}. However, performing full Bayesian (or
even simple Fisher matrix) parameter estimation calculations for EMRIs
remains particularly challenging because of the extreme accuracy
required in GW flux calculations. We plan to address this in a
follow-up study. The BH solutions we derived can be extended to
density-pressure profiles not necessarily associated with DM halos,
such as accretion disks. Finally, recent studies have emphasized the
importance of the BH spin in environmental effects on GW
signals~\cite{Dyson:2025dlj}. As a natural extension of this work, we
will study EMRIs evolving in a Kerr BH background.

\begin{figure*}[!ht]
\centering
\includegraphics[width=0.95\textwidth]{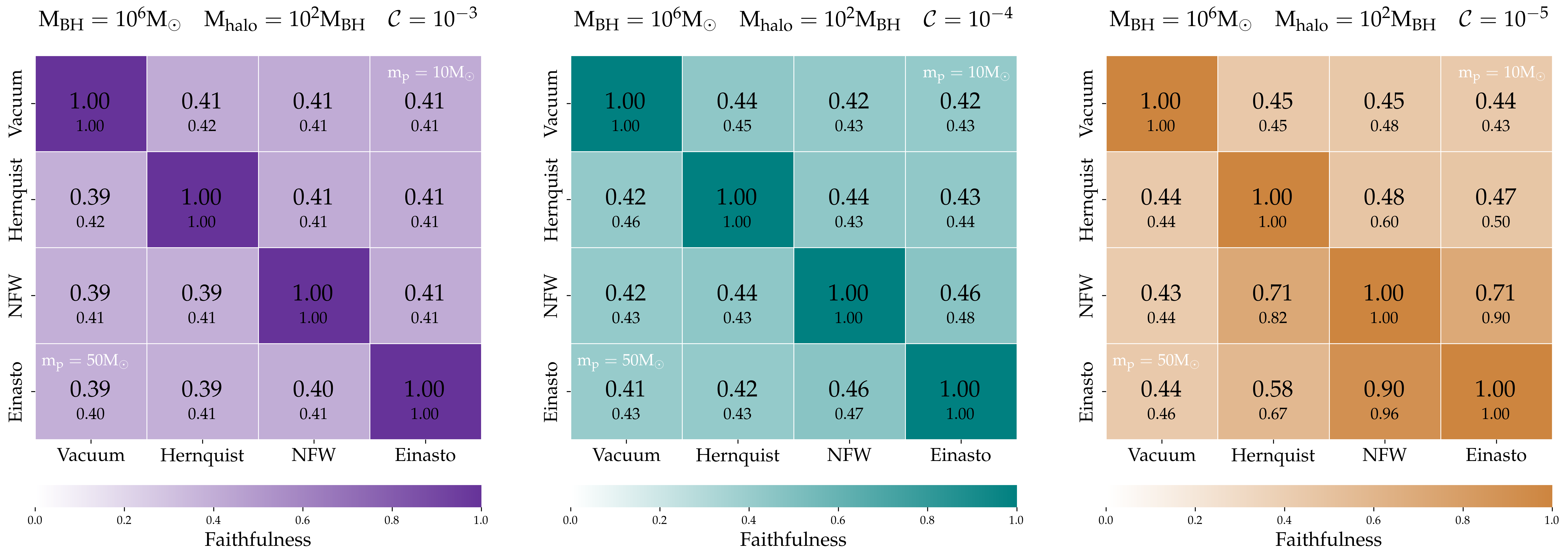} \\  
\includegraphics[width=0.95\textwidth]{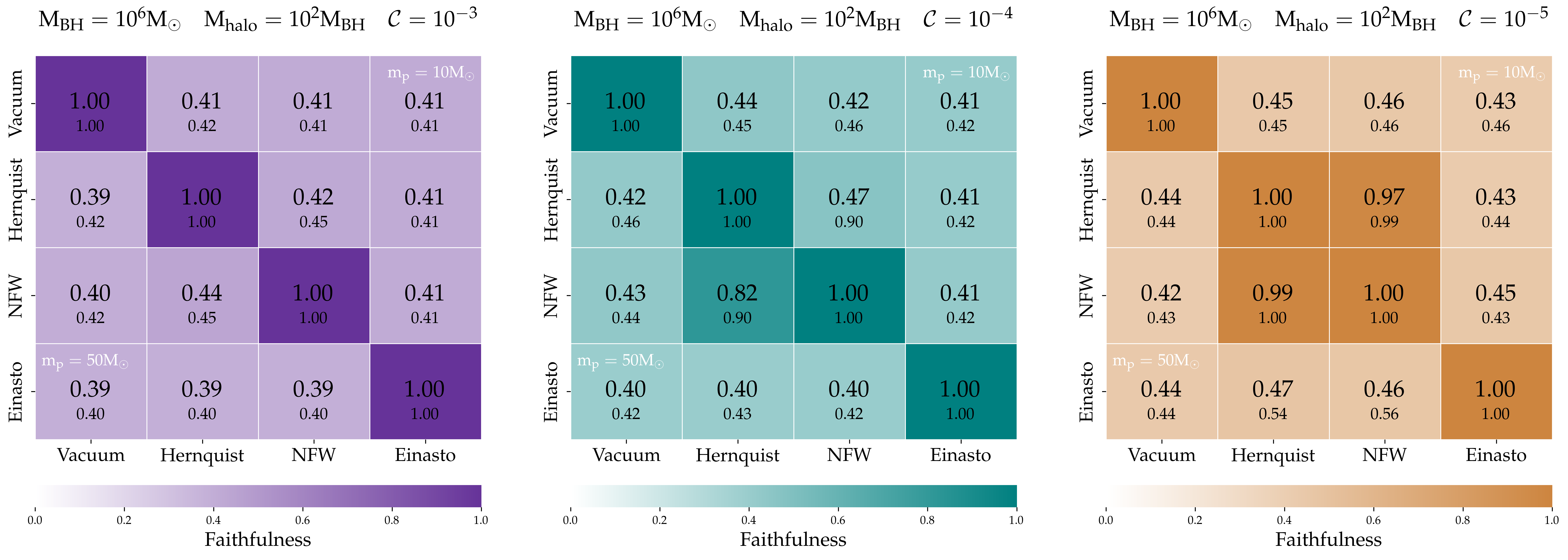}    
\caption{Faithfulness density matrices for different DM profiles
  (Hernquist, NFW, and Einasto) compared to the vacuum case and among
  each other. The three panels correspond to halo compactness values
  of \( \mathcal{C} =(10^{-3}, 10^{-4}, 10^{-5}) \), assuming a halo
  mass of \( M_{\text{halo}} = 10^{2} M_{\text{BH}} \).  The upper and
  lower diagonals of each matrix display faithfulness values computed
  for systems with $(M_{\text{BH}}, m_p) = (10^6,10) M_\odot$ and
  $(M_{\text{BH}}, m_p) = (10^6,50) M_\odot$, respectively.  Large
  numbers in each cell indicate faithfulness values computed for a
  one-year evolution, while smaller numbers correspond to results
  obtained for a six-month evolution.  At variance with the results
  shown in Fig.~\ref{fig:faithfulness_Mh100}, in the first and second
  row we assume $(r_c = a_0, r_e=a_0)$ and $(r_c = 5 a_0, r_e=a_0/2)$
  for the NFW and the Einasto profiles, respectively.}
\label{fig:different_parameters}
\end{figure*}

\begin{acknowledgments}
We would thank the anonymous 
Referee for their valuable comments 
which have improved the quality of 
our manuscript. 
N.S. and E.B. are supported by NSF Grants No. AST-2307146, PHY-2207502, PHY-090003 and PHY-20043, by 
NASA Grants No. 20-LPS20-0011 and 21-ATP21-0010, by the John Templeton Foundation Grant 62840, by the 
Simons Foundation. A.M. and S.G. acknowledge financial support from MUR PRIN Grants No.~2022-Z9X4XS and 
No.~2020KB33TP.
The Authors  acknowledge support from the ITA-USA Science and Technology Cooperation program, supported 
by the Ministry of Foreign Affairs of Italy (MAECI), grant No.~PGR01167. This work was carried out at the Advanced 
Research Computing at Hopkins (ARCH) core facility (\url{https://www.arch.jhu.edu/}), which is supported by the NSF 
Grant No.~OAC-1920103. 
\end{acknowledgments}

\begin{table*}
\begin{tabular}{p{0.05\textwidth} p{0.05\textwidth} p{0.05\textwidth} p{0.07\textwidth} p{0.07\textwidth} p{0.07\textwidth} p{0.07\textwidth} p{0.07\textwidth} p{0.07\textwidth}}
\hline
\hline
{\small density} &  $m_p$ & {\small T} & $\Delta \phi$ & $\Delta \phi $ & $\Delta \phi$ & $\mathcal{F}$ & $\mathcal{F} $ & $\mathcal{F}$ \\
 {\small profile} &  $[M_\odot]$ & [year]  & $\ell_{\rm max}=2$ & $\ell_{\rm max}=3$ & $\ell_{\rm max}=4$ & $\ell_{\rm max}=2$ & $\ell_{\rm max}=3$ & $ \ell_{\rm max}=4$ \\
\hline
\hline
{\small Hernquist} & 10 & 1 & 436 & 416 & 412 & 0.41 & 0.41  & 0.41\\
                  &    & 0.5 & 180 & 170 & 169 & 0.42 & 0.42 & 0.41  \\
                  & 50 & 1 & 269 & 256 & 255 & 0.39 & 0.40 & 0.40 \\
                  &    & 0.5 & 100 & 96& 96 & 0.42 & 0.41 & 0.41  \\
\hline
{\small NFW} & 10 & 1 & 386 & 367 & 364 & 0.41 & 0.41 & 0.42 \\
             &    & 0.5 & 159 & 151 & 150 & 0.41 & 0.41  & 0.42 \\
             & 50 & 1 & 237 & 226 & 225 & 0.40 & 0.40 & 0.40 \\
             &    & 0.5 & 89 & 85 & 84 & 0.41 & 0.42 & 0.41 \\
\hline
{\small Einasto} & 10 & 1 & 1360 & 1310 & 1303 & 0.41 & 0.41 & 0.41 \\
                 &    & 0.5 & 557 & 535 & 532 & 0.41 & 0.41  & 0.41 \\
                 & 50 & 1 & 846 & 816 & 812 & 0.39 & 0.39 & 0.39 \\
                 &    & 0.5 & 320 & 309 & 308 & 0.40 & 0.40 & 0.40 \\
\hline
\hline
\end{tabular}
\caption{GW dephasing $\Delta\Phi$ and faithfulness ${\cal F}$
  with respect to vacuum computed for EMRIs with a primary of $M=10^6M_\odot$ and a different secondary masses
  ($m_p=10 M_\odot$ or $m_p=50M_\odot$), evolving for either one year or six months in different DM profiles. We assume halo masses of 
  $\rm{M}_{\rm{halo}} = 10^3 \rm{M}_\odot$
  and a compactness of ${\cal C}=10^{-3}$. 
  The dephasing and faithfulness are computed
  using the GW fluxes in Eq.~\eqref{eq:energyfluxes}, truncating the
  sum over $\ell$ at $\ell=\ell_{\rm max}$.  The values for
  $\ell_{\rm max}=2$ correspond to the results discussed in
  Sec.~\ref{sec:results}.}
  \label{tab:lmodes}
\end{table*}

\appendix

%
\section{Dependence on $\ell_\text{max}$}\label{app:lmax}
%

In the main text we computed GW fluxes fixing $\ell_{\rm max}$ in
Eq.~\eqref{eq:energyfluxes} to the leading ($\ell=2$) contribution. In
this appendix we assess the impact of higher-order multipoles in the
calculations of the dephasing and faithfulness. We compare the results
presented in Sec.~\ref{sec:results} with the values of $\Delta \Phi$
and ${\cal F}$ obtained by including higher multipoles
($\ell_\mathrm{max} = 2,3,4$) across all three density profiles and
for different values of the secondary mass. We focus on EMRIs with
halo mass $\mathrm{M}_{\mathrm{halo}} = 10^2 M_\mathrm{BH}$ and
compactness $\mathcal{C} = 10^{-3}$, as these parameters yield the
largest deviations from the vacuum scenario. The results are presented
in Table~\ref{tab:lmodes}.  While increasing $\ell_{\rm max}$ reduces the dephasing by $\sim 5\%$ relative to the baseline
configuration, this effect tends to saturate for all halo models and EMRI configurations. More importantly, the
distinguishability between signals in vacuum and matter environments remains unaffected. The faithfulness hardly changes at all when we consider different values of $\ell_{\rm max}$. This confirms the
robustness of our main findings.

%
\section{The effect of halo cutoff parameters on the DM detectability}\label{app:differentprofiles}
%

To assess the impact of the NFW cutoff radius and of $r_e$ in Eq.~\eqref{eq:einsteineq} for the Einasto model, we compute the faithfulness 
for the same EMRI configurations shown in Fig.~\ref{fig:faithfulness_Mh100}, changing 
the values of $(r_c,r_e)$ with respect 
to the baseline values used in Sec.~\ref{sec:results}, i.e.
($r_c = 5a_0,r_e = a_0$). In the first row of Fig.~\ref{fig:different_parameters}, we reduce the NFW cutoff radius to 
$r_c = a_0$ while keeping $r_e =a_0$. In the second row, we decrease the Einasto parameter to $r_e = a_0/2$ while keeping $r_c = 5_a0$.
The resulting values of \( {\cal F} \), whether computed against vacuum templates or
between different DM models, remain consistent with those in
Fig.~\ref{fig:faithfulness_Mh100}, where we assumed for all binaries $(r_c=5a_0, r_e=a_0)$. This confirms that the distinguishability of
halo signals is largely insensitive to the mass and scale of the
distribution, while it primarily depends on the overall halo
compactness.
\bibliography{biblio}

\end{document}